\documentclass[aps,prd,unsortedaddress,superscriptaddress,showpacs,twocolumn, floatfix]{revtex4}

\usepackage{amsmath}
\usepackage{graphicx}

\begin{document}

\title{A Precision Measurement of the Muon Decay Parameters $\rho$ and
  $\delta$}

\affiliation{University of Alberta, Edmonton, AB, T6G 2J1, Canada}
\affiliation{University of British Columbia, Vancouver, BC, V6T 1Z1, Canada}
\affiliation{Kurchatov Institute, Moscow, 123182, Russia}
\affiliation{University of Montreal, Montreal, QC, H3C 3J7, Canada}
\affiliation{University of Regina, Regina, SK, S4S 0A2, Canada}
\affiliation{Texas A\&M University, College Station, TX 77843, U.S.A.}
\affiliation{TRIUMF, Vancouver, BC, V6T 2A3, Canada}
\affiliation{Valparaiso University, Valparaiso, IN 46383, U.S.A.}

\author{R.P.~MacDonald}
\affiliation{University of Alberta, Edmonton, AB, T6G 2J1, Canada}

\author{R.~Bayes}
\altaffiliation[Affiliated with: ]{Univ.~of Victoria,
Victoria, BC.}
\affiliation{TRIUMF, Vancouver, BC, V6T 2A3, Canada}

\author{J.~Bueno}
\affiliation{University of British Columbia, Vancouver, BC, V6T 1Z1, Canada}

\author{Yu.I.~Davydov}
\altaffiliation[Present address: ]{JINR, Dubna, Russia}
\affiliation{TRIUMF, Vancouver, BC, V6T 2A3, Canada}

\author{P.~Depommier}
\affiliation{University of Montreal, Montreal, QC, H3C 3J7, Canada}

\author{W.~Faszer}
\affiliation{TRIUMF, Vancouver, BC, V6T 2A3, Canada}

\author{M.C.~Fujiwara}
\affiliation{TRIUMF, Vancouver, BC, V6T 2A3, Canada}

\author{C.A.~Gagliardi}
\affiliation{Texas A\&M University, College Station, TX 77843, U.S.A.}

\author{A.~Gaponenko}
\altaffiliation[Present address: ]{LBNL,
Berkeley, CA.}
\affiliation{University of Alberta, Edmonton, AB, T6G 2J1, Canada}

\author{D.R.~Gill}
\affiliation{TRIUMF, Vancouver, BC, V6T 2A3, Canada}

\author{A.~Grossheim}
\affiliation{TRIUMF, Vancouver, BC, V6T 2A3, Canada}

\author{P.~Gumplinger}
\affiliation{TRIUMF, Vancouver, BC, V6T 2A3, Canada}

\author{M.D.~Hasinoff}
\affiliation{University of British Columbia, Vancouver, BC, V6T 1Z1, Canada}

\author{R.S.~Henderson}
\affiliation{TRIUMF, Vancouver, BC, V6T 2A3, Canada}

\author{A.~Hillairet}
\altaffiliation[Affiliated with: ]{Univ.~of Victoria,
Victoria, BC.}
\affiliation{TRIUMF, Vancouver, BC, V6T 2A3, Canada}

\author{J.~Hu}
\affiliation{TRIUMF, Vancouver, BC, V6T 2A3, Canada}

\author{B.~Jamieson}
\affiliation{University of British Columbia, Vancouver, BC, V6T 1Z1, Canada}

\author{P.~Kitching}
\affiliation{TRIUMF, Vancouver, BC, V6T 2A3, Canada}

\author{D.D.~Koetke}
\affiliation{Valparaiso University, Valparaiso, IN 46383, U.S.A.}

\author{G.M.~Marshall}
\affiliation{TRIUMF, Vancouver, BC, V6T 2A3, Canada}

\author{E.L.~Mathie}
\affiliation{University of Regina, Regina, SK, S4S 0A2, Canada}

\author{R.E.~Mischke}
\affiliation{TRIUMF, Vancouver, BC, V6T 2A3, Canada}

\author{J.R.~Musser}
\altaffiliation[Present address: ]{Arkansas Tech University,
Russellville, AR.}
\affiliation{Texas A\&M University, College Station, TX 77843, U.S.A.}

\author{M.~Nozar}
\affiliation{TRIUMF, Vancouver, BC, V6T 2A3, Canada}

\author{K.~Olchanski}
\affiliation{TRIUMF, Vancouver, BC, V6T 2A3, Canada}

\author{A.~Olin}
\altaffiliation[Affiliated with: ]{Univ.~of Victoria,
Victoria, BC.}
\affiliation{TRIUMF, Vancouver, BC, V6T 2A3, Canada}

\author{R.~Openshaw}
\affiliation{TRIUMF, Vancouver, BC, V6T 2A3, Canada}

\author{J.-M.~Poutissou}
\affiliation{TRIUMF, Vancouver, BC, V6T 2A3, Canada}

\author{R.~Poutissou}
\affiliation{TRIUMF, Vancouver, BC, V6T 2A3, Canada}

\author{M.A.~Quraan}
\altaffiliation[Present address: ]{VSM Medtech Ltd., Coquitlam, BC.}
\affiliation{University of Alberta, Edmonton, AB, T6G 2J1, Canada}

\author{V.~Selivanov}
\affiliation{Kurchatov Institute, Moscow, 123182, Russia}

\author{G.~Sheffer}
\affiliation{TRIUMF, Vancouver, BC, V6T 2A3, Canada}

\author{B.~Shin}
\altaffiliation[Affiliated with: ]{Univ.~of Saskatchewan,
Saskatoon, SK.}
\affiliation{TRIUMF, Vancouver, BC, V6T 2A3, Canada}

\author{T.D.S.~Stanislaus}
\affiliation{Valparaiso University, Valparaiso, IN 46383, U.S.A.}

\author{R.~Tacik}
\affiliation{University of Regina, Regina, SK, S4S 0A2, Canada}

\author{R.E.~Tribble}
\affiliation{Texas A\&M University, College Station, TX 77843, U.S.A.}

\collaboration{TWIST Collaboration}
\noaffiliation

\date{\today}

\begin{abstract}

  The TWIST collaboration has performed new measurements of two of the
  parameters that describe muon decay: $\rho$, which governs the shape
  of the overall momentum spectrum, and $\delta$, which governs the
  momentum dependence of the parity-violating decay asymmetry.  This
  analysis gives the results $\rho=0.75014\pm 0.00017(\text{stat})\pm
  0.00044(\text{syst})\pm 0.00011(\eta)$, where the last uncertainty
  arises from the correlation between $\rho$ and the decay parameter
  $\eta$, and $\delta = 0.75067\pm 0.00030(\text{stat})\pm
  0.00067(\text{syst})$.  These are consistent with the value of $3/4$
  given for both parameters in the Standard Model of particle physics,
  and are a factor of two more precise than the measurements
  previously published by TWIST.  A new global analysis of all available muon decay data incorporating these results is presented.
  Improved lower and upper limits on the decay parameter
  $P_\mu^\pi\xi$ of $0.99524 < P_\mu^\pi\xi \leq \xi < 1.00091$ at
  90\% confidence are determined, where $P_\mu^\pi$ is the
  polarization of the muon when it is created during pion decay, and
  $\xi$ governs the muon decay asymmetry.   These results
  set new model-independent constraints on the possible weak
  interactions of right-handed particles.  Specific implications for
  left-right symmetric models are discussed.

\end{abstract}

\pacs{13.35.Bv, 14.60.Ef, 12.60.Cn}% PACS codes

\maketitle

% typing shortcuts
\newcommand{\e}[1]{\ensuremath{\times 10^{#1}}}
\newcommand{\us}{\ensuremath{\mu}s}      % Microseconds.
\newcommand{\um}{\mbox{$\mu$m}}          % Microns
\newcommand{\kevc}{keV/$c$}
\newcommand{\mevc}{MeV/$c$}
\newcommand{\mevcc}{MeV/$c^2$}
\newcommand{\pmuxi}{\ensuremath{P_\mu \xi}}
\newcommand{\pmupixi}{\ensuremath{P_\mu^\pi \xi}}
\newcommand{\abs}[1]{\ensuremath{\left\lvert#1\right\rvert}} % Absolute value

\section{Introduction}
\label{sec:intro}

% Motivation and theory.  

In the Standard Model (SM) of particle physics, the charged-current
weak interaction violates parity maximally---only left-handed
particles (or right-handed antiparticles) are affected.  The TWIST
experiment is a high-precision search for contributions from non-SM
forms of the charged-current weak interaction, including
parity-conserving currents.

The decay of the positive muon into a positron and two neutrinos,
{$\mu^+ \rightarrow e^+ \nu_e \bar \nu_\mu$}, is a purely leptonic
process, making it an excellent system for high-precision studies of
the weak interaction.  It proceeds through the charged weak
current---mediated by the $W$ boson---and can be described to a good
approximation as a four-fermion point interaction.  The matrix element
for the most general Lorentz-invariant, local, four-fermion
description of muon decay can then be written as
\begin{equation}
  \label{eq:mudecay_matrixelem}
  M = \frac{4 G_F}{\sqrt{2}}
  \sum_{\substack{\gamma=S,V,T\\\epsilon,\mu=L,R}} g_{\epsilon\mu}^\gamma 
  \bigl\langle\bar\psi_{e_\epsilon} \bigl\vert\Gamma^\gamma\bigr\vert
  \psi_{\nu_e}\bigr\rangle
  \bigl\langle\bar\psi_{\nu_\mu}
  \bigl\vert\Gamma_\gamma\bigr\vert \psi_{\mu_\mu}\bigr\rangle,
\end{equation}
where the $g_{\epsilon\mu}^\gamma$ specify the scalar, vector, and
tensor couplings between $\mu$-handed muons and $\epsilon$-handed
positrons~\cite{Fetscher86:muon_decay}, and satisfy certain
normalizations and constraints.  In the Standard Model, $g_{LL}^V=1$
and all other coupling constants are zero.  The probability
$Q_{\epsilon\mu}$ $(\epsilon,\mu = L,R)$ for the decay of a
$\mu$-handed muon into an $\epsilon$-handed positron is given by
\begin{equation}
  \label{eq:Qem}
  Q_{\epsilon \mu} = \frac{1}{4} \abs{g_{\epsilon\mu}^{S}}^{2} +
  \abs{g_{\epsilon\mu}^{V}}^{2} +
  3(1-\delta_{\epsilon\mu}) \abs{g_{\epsilon\mu}^{T}}^{2},
\end{equation}
where $\delta_{\epsilon\mu}=1$ for $\epsilon=\mu$ and
$\delta_{\epsilon\mu}=0$ for $\epsilon\neq\mu$.  The probability:
\begin{align}
    Q_R^\mu & =  \frac{1}{4} \abs{g^S_{LR}}^2 + \frac{1}{4} \abs{g^S_{RR}}^2
    + \abs{g^V_{LR}}^2 + \abs{g^V_{RR}}^2 + 3 \abs{g^T_{LR}}^2
 \label{eq:QmuR}    
\end{align}
sets a model independent limit on any muon right-handed couplings
%\cite{Fetscher86:muon_decay,PDBook2006}.
\cite{Fetscher86:muon_decay,PDBook2006:mudecay}.

The differential muon decay spectrum~\cite{michel50:_michel_param,
  Bouchiat-Michel, Kinoshita-Sirlin} can be described following the
notation of Fetscher and Gerber~\cite{PDBook2006} as
\begin{equation}
  \label{eq:mudecay_michel}
  \begin{split}
    \frac{d^2\Gamma}{dx\, d(\cos\theta_s)} = & \frac{m_\mu}{4\pi^3}
    E_{max}^4 G_F^2 \sqrt{x^2 - x_0^2} \\
    & {} \times \left( F_{IS}(x) + P_\mu\xi\cos\theta_s F_{AS}(x)
    \right) 
  \end{split}
\end{equation}
where $G_F$ is the Fermi coupling constant, $\theta_s$ is the angle
between the muon spin and the positron momentum, $E_{max}\approx
52.8$~MeV is the kinematic maximum positron energy, $x = E_e/E_{max}$
is the positron's reduced energy, $x_0 = m_e/E_{max}$ is the minimum
possible value of $x$, corresponding to a positron at rest, and
$P_\mu$ is the degree of muon polarization at the time of decay.
$P_\mu$ can be used to determine $P_\mu^\pi$, the degree of muon
polarization at its creation from pion decay, when the amount of
depolarization undergone by the muon is known.  The two components of
the decay spectrum in Eqn.~\eqref{eq:mudecay_michel} are the isotropic
component:
\begin{equation}
  \label{eq:Fis}
  \begin{split}
    F_{IS}(x) =&\, x(1-x) + \rho\,\frac{2}{9} ( 4x^2 - 3x - x_0^2 ) \\
    & {} + \eta\, x_0 (1-x) + \text{R.C.}
  \end{split}
\end{equation}
and the anisotropic component:
\begin{equation}
  \label{eq:Fas}
  \begin{split}
    F_{AS}(x) =&\, \frac{1}{3} \sqrt{x^2 - x_0^2} \bigg\{ 1 - x \\
    & {} + \frac{2}{3}\delta\,\bigg[ 4x - 3 +
    \left(\sqrt{1-x_0^2}-1\right) \bigg] \bigg\} + \text{R.C.}
  \end{split}
\end{equation}
R.C.\ represents the electromagnetic radiative corrections, which have
been calculated to $O(\alpha^2)$.  Corrections due to the strong
interaction in loops give a fractional contribution on the order of
$4\e{-7}$~\cite{davydychev01:hadronic_mudecay}, which is more than two
orders of magnitude smaller than the ultimate precision goals of
TWIST.  The quantities $\rho$, $\delta$, $\eta$, and $\xi$, often
called the Michel parameters, are bilinear combinations of the weak
coupling constants and describe the shape of the decay spectrum.
These decay parameters can be used in combination with other muon
decay measurements, such as inverse muon decay ({$e^-\nu_\mu
  \rightarrow \mu^- \nu_e$}) and the polarization of the decay
positron, to determine limits on the weak coupling constants.

 Left-Right Symmetric (LRS) models~\cite{HerczegLRS:1986} comprise an interesting class of extensions to the SM.
These models include a right-handed weak coupling, which is suppressed
by the mass of the associated gauge boson.  LRS models contain four
charged gauge bosons ($W_1^\pm$, $W_2^\pm$) with masses $m_1$ and
$m_2$, and two additional massive neutral gauge bosons.  The mass
eigenstates $W_1$ and $W_2$ are related to the weak eigenstates $W_L$
and $W_R$ through a mixing angle $\zeta$.  $g_L$ and $g_R$ are the
coupling strengths of the LRS weak interaction to left- and
right-handed particles.  LRS models affect the muon decay spectrum,
including a modification to the decay parameter $\rho$:
\begin{equation}
  \label{eq:lrs_rho}
  \rho \simeq \frac{3}{4} \left[ 
    1 - 2 \left(
      \frac{g_R}{g_L} \zeta
    \right)^2
  \right],
\end{equation}
therefore requiring $\rho\leq 0.75$.  To a good approximation, the
value of $\delta$ is unaffected by LRS models.

Many other proposed SM extensions also lead to modifications of the Michel parameters.  For example, supersymmetric models can lead to a non-zero value for $g^S_{RR}$ \cite{Profumo07}.  The values of $\xi$ and $\delta$ can be combined to provide a model-independent limit on right-handed couplings in muon decay:
\begin{equation}
Q^{\mu}_R = \frac{1}{2} \left[ 1 + \frac{1}{3} \xi - \frac{16}{9} \xi \delta \right] .
\end{equation}

Previous measurements by TWIST had set new limits, of $\rho=0.75080\pm
0.00032(\text{stat.})\pm 0.00097(\text{syst.})\pm
0.00023(\eta)$~\cite{musser:2005} and $\delta=0.74964\pm
0.00066(\text{stat.})\pm 0.00112(\text{syst.})$~\cite{gaponenko:2005}.
This paper presents new results from a refined analysis of newer data,
providing details about the TWIST experiment and analysis
(Sects.~\ref{sec:experiment} and~\ref{sec:analysis}) including
improvements over the previous studies (Sect.~\ref{sec:improvements}),
discusses the validation of the Monte Carlo simulations
(Sect.~\ref{sec:usstops}), describes the current state of the
systematic uncertainties in the experiment
(Sect.~\ref{sec:systematics}), and presents new measurements of $\rho$
and $\delta$ (Sect.~\ref{sec:results}).  These new results are then
incorporated into a global analysis of all muon decay data
(Sect.~\ref{sec:globalanalysis}).

\section{Experimental Procedures}
\label{sec:experiment}

% Experimental Setup & hardware; Data Sets; Simulation

\subsection{Experimental Setup}
\label{sec:setup}

Highly polarized muons resulted from the decay of pions stopping at
the surface of a carbon production target bombarded by 500~MeV protons
at TRIUMF.  The M13 beamline~\cite{oram81:commis_m13_triumf} selected
positive muons with a momentum of 29.6~\mevc, with a momentum bite of
$\Delta p/p\approx 1$\% FWHM, and delivered these in vacuum to the
TWIST spectrometer~\cite{TWIST_DCs:2005} at a typical rate of 2.5\e{3}
per second.  The beam had a contamination of positrons, at the same
momentum, with a typical rate of 22~kHz, as well as a small fraction
of pions.

The muon beam was characterized and tuned using a low pressure (8~kPa
dimethylether (DME) gas) removable beam monitoring chamber located
upstream of the TWIST spectrometer~\cite{TEC2006}.  The beam monitor
was inserted for measurement of the beam properties, and removed
during data-taking.

\begin{figure}[!hbt]
    \includegraphics[width=3.4in]{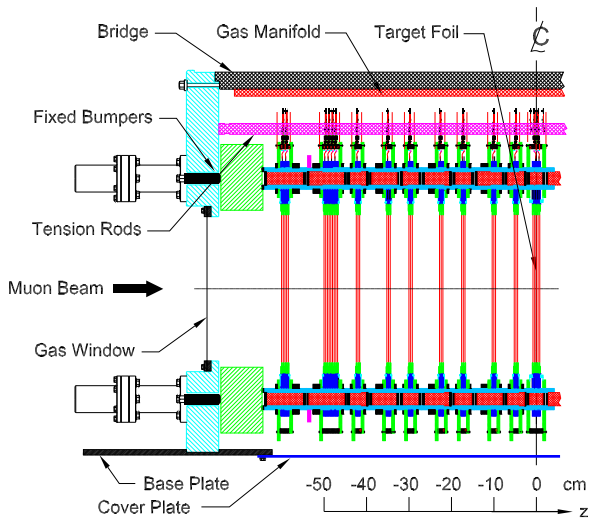}
    \caption{(Color online.)  Side view of the upstream half of the
      TWIST spectrometer planar chambers and support structures.
      Muons stopped in the target foil, which also served as a chamber
      cathode.  The use of precision spacers and tension control
      ensured that the chamber wire positions were known to 30~\um\ in
      $z$.  The spectrometer is symmetric about the target foil.}
  \label{fig:halfstack}
\end{figure}

The TWIST spectrometer~\cite{TWIST_DCs:2005} was designed to measure a
broad range of the muon decay spectrum, allowing the simultaneous
determination of the decay parameters.  The detector consisted of 44
drift chambers (DCs) and 12 multiwire proportional chambers (PCs) in a
planar geometry, symmetric about a muon stopping target foil at the
centre.  Figure~\ref{fig:halfstack} shows the upstream half of the
detector and the four PCs surrounding the stopping target.  The
chambers were placed in a highly uniform 2~T solenoidal magnetic
field.  The $z$ axis was defined to be the detector axis.

To reduce scattering and to allow the muons to reach the central
target foil, the detector was designed to be very thin; there was
approximately 140~mg/cm$^2$ of material from the vacuum of the M13
beamline through to the centre of the stopping target.

The muon stopping target was a $71\pm 1$~\um\ foil of 99.999\% pure
aluminum, which also served as a cathode foil for the adjacent two PCs
(see below).  The muon stopping distribution, as determined from the
last DC or PC plane in which the muon left a signal, was used in a
feedback loop to control the fractions of He and CO$_2$ in a gas
degrader at the end of the beamline, in order to maintain the average
stopping position of the selected muons at the centre of the target.
Muons were required to be recorded by the PC immediately before the
stopping target (PC~6) and not by the PC immediately after the target
(PC~7); simulations showed that $97.0\pm 1.5$\% of selected muons
stopped in the target, with the rest stopping in the CF$_4$/isobutane,
the cathode foils, or the wires in the vicinity of the target.  The experiment watched
for muons decaying at rest into positrons; decays of muons in flight
were identified and discarded.

Each PC and DC consisted of a wire plane and two cathode foil planes,
all oriented perpendicular to the detector axis.  Chambers were
grouped together in modules as described below, and the cathode foils
interior to each module were shared between adjacent chambers.  Each
wire plane was in either $U$ or $V$ orientation, and these were at
right angles to each other and $45^{\circ}$ to the vertical.

Four PCs were located at each end of the detector, and two on either
side of the stopping target.  Each PC included 160 sense wires at 2~mm
pitch.  A fast drift gas (CF$_4$/isobutane) was used in these
chambers.  The PCs mainly provided particle timing information; in
addition the widths of the timing signals from the PCs were used to
discriminate muons from positrons.

Each half of the detector included 22 DCs, with a slow drift gas
(DME), which had a small Lorentz angle; these provided precise
measurements of the $e^+$ position as it crossed the chamber.  Each DC
plane~\cite{DavydovDCs:2001} included 80 sense wires at 4~mm pitch.
The DC planes were very slightly asymmetric in the direction along the detector
axis, in that the wires were located 150~\um\ off of the centre of the
cell.  28 of the DCs were paired together in 14 modules, with $U$ and
$V$ wire planes in each pair.  Additional ``dense'' stacks of eight
planes each were located near either end of the detector.

The space between the chambers was filled with a 3\%~N$_2$ and 97\%~He
gas, minimizing the material thickness of the detector.  The chamber
gas and the helium mixture were maintained at atmospheric pressure.
The differential pressure between the chambers and the surrounding
volumes was controlled to stabilize the positions of the chamber
foils.

Chamber alignments perpendicular to the detector axis were measured
using 120 MeV/c pion tracks with the magnetic field off.  Alignment
measurements were taken at the beginning and end of the data-taking
period.  These data were also used to measure wire time offsets
introduced by the electronics and cabling.  The plane positions and
rotations about the beam direction were determined to an accuracy of 10~\um\ and 0.4~mrad.
Relative wire positions were known to 3~\um.  The alignment of the
chambers to the magnetic field was measured using positron tracks in
the 2~T field, and had an uncertainty of 0.03~mrad.

The chambers were positioned within the 1~m bore of a
liquid-helium-cooled superconducting solenoid.  The magnet was placed
inside a cube-shaped steel yoke, approximately 3~m per side, designed
to increase the uniformity of the field within the detector region.
The shape of the $z$ component of the magnetic field was mapped using
an array of Hall probes mounted on a radial arm, aligned along the
solenoid axis and calibrated using an NMR probe.  The probes'
positions at each field sample were known to $\pm 1$~mm, and the field
was mapped to a precision of $\pm 1\e{-4}$~T.  The standard operating
setting for TWIST is 2~T at the centre of the solenoid; at this
setting the field is uniform to within 8\e{-3}~T within the tracking
region.  A finite-element simulation of the magnetic field was
performed using \textsc{Opera3d}~\cite{opera3d}, providing the full
three-dimensional magnetic field throughout the detector volume and
well outside the measured region.  Within the tracking region the 
 simulated field map agreed with
the measurements to within $\pm 2\e{-4}$~T.

Event acquisition was triggered by a thin scintillator upstream of the
spectrometer.  Only 1.8\% of events were triggered by a beam positron,
in spite of the much higher rate of beam positrons compared to muons;
most beam positrons were discriminated against by the trigger
threshold.  Remaining beam positrons were identified with high
efficiency in the data since they left signals in the full length of
the detector.

Preamplifier chips were mounted directly on the chambers and drove
custom postamplifier and discriminator modules placed 2~m away.  The
chamber behavior was exceptionally stable, with less than one high
voltage trip per month.  All output channels were functional.

Data acquisition from scintillators and tracking
chambers~\cite{PoutissouDAQ:2003} was performed using LeCroy
Model~1877 TDCs.  The trigger and TDC read-out recorded signals in
0.5~ns time bins from 6~\us\ before to 10~\us\ after a muon passed
through the trigger scintillator.  The start and stop of each pulse
was recorded so that the pulse width could be determined; these widths
can be related to the amount of energy deposited in the cell.  In this
configuration up to eight wire signals could be recorded for each wire
in any triggered event.  A fixed blanking time of 80~\us\ was imposed
after each accepted trigger, to allow each TDC to finish conversion
before the next event was recorded.

The data considered for this analysis were more than 1.5\e{9} muon
decays, taken during 2004.  The data sets, summarized in
Table~\ref{tab:datasets}, were taken under a variety of conditions
(low polarization from beam steering, rate, muon stopping position, etc.).  Provided the
simulation reproduced these conditions correctly, the decay parameters
extracted from the data should be independent of the run conditions.
\begin{table}[!hbt]
  \caption{Description of data sets, in chronological order.}
  \label{tab:datasets}
  \begin{tabular}{lcc}
    \hline
    \hline
    Description & Events & Accepted \\
    & (\e{6}) & (\e{6}) \\
    \hline
    Low Polarization, Centred  & $209$ & $8$\\
    Low Polarization, PC5 Stops & $94$  & $2$\\
    Centred Stops              & $287$ & $11$\\
    3/4 Stops A                 & $323$ & $12$\\
    High Rate                   & $198$ & $7$\\
    Aperture                    & $263$ & $9$\\
    3/4 Stops B                 & $157$ & $6$\\
    \hline
    Total                       & $1531$ & $55$ \\
    \hline
    \hline
  \end{tabular}
\end{table}

\section{Simulation}
\label{sec:simulation}

The decay parameters were extracted from the data by comparing the
data to a simulated spectrum, as discussed in Sect.~\ref{sec:McFit}.
Monte Carlo (MC) simulations were run to match the conditions of each
of the seven main data sets, in addition to simulations run for
studying systematic uncertainties.  Each
simulation included 2--3 times as many muon decays as the
corresponding data set.  The simulations were implemented using
Geant~3.21.  The output was in exactly the same format as the files
produced by the data acquisition system, and were processed in the
same way as real data.  Space-time relationships (STR) determined with
\textsc{garfield}~\cite{garfield} were used to model drift chamber
response.

The theoretical decay spectrum included full radiative corrections at
$O(\alpha)$~\cite{arbuzov02:treeRCs}, as well as $O(\alpha^2L^2)$ and
$O(\alpha^2L)$~\cite{arbuzov02:a2L2RCs,arbuzov02:a2LRCs}, where
$L=\log(m_\mu^2/m_e^2) \approx 10.66$.  $O(\alpha^2L^0)$ terms have
also been calculated~\cite{anastasiou:a2RCs}; the effect of neglecting
these last terms has been evaluated and shown to be negligible (see
Sect.~\ref{sec:othersystematics}).  All radiative corrections are
calculated within the Standard Model.  The values of the muon decay
parameters assumed by the simulation in its theoretical decay spectrum
were kept hidden until the end of the study; see Sect.~\ref{sec:McFit}
below for details.

The simulation included all known depolarization effects, including
interactions with the magnetic field as the muon enters the solenoid
and depolarization in the stopping target.

\section{Data Analysis}
\label{sec:analysis}

% Analysis Flow; Event Reconstruction; Spectrum Fitting

The full analysis procedure was applied in the same way to both data
and simulation.  To the level that the simulation accurately
represented the data, this canceled spectrum distortions due to
detector response, positron energy loss and scattering, reconstruction
biases, and other effects, which would otherwise lead to systematic
errors in the measurement.

Due to the large amount of simulation and analysis required, the
Western Canada Research Grid (WestGrid) was used.  TWIST used
approximately 10,000 CPU-days for this simulation and analysis.

\subsection{Event Reconstruction}
\label{sec:reconstruction}

To reconstruct an event, the hits---signal times on individual
wires---were first grouped based on timing information from the PCs.
Tracks within these groups were then identified using the distribution
of DC hits in space and time, as well as the hit widths.  DC hits
associated with decay positrons were then used to reconstruct the
energy and angle of each positron.

Pattern recognition was performed on the decay positrons using the
spatial hit distributions, to determine an initial estimate of the
positron track.  The track fit parameters were the position and momentum
three-vectors at the  DC closest to the target; the
main parameters of interest were the positron's total momentum $p$,
and the angle $\theta$ between the positron momentum vector and the
detector axis.  Multiple overlapping tracks could be distinguished at
this stage.  The initial track estimate was then refined in a $\chi^2$
fit using the hits' drift times, in combination with maps of the STRs
of the drift chamber cell as provided by the \textsc{garfield} chamber
simulation software~\cite{garfield}.  The positron track was assumed
to deviate from a helix by continuous energy loss through the gas
volumes and discrete energy loss through each foil encountered, using
mean energy loss formulas~\cite{PDBook2006}.  Tracks were allowed
discrete deflections at each DC module and in the dense stacks.  The
deflection angles were fit parameters, with associated penalties to
the fit $\chi^2$, based on the method described by
Lutz~\cite{lutz88:kinks}.

Event and track selection cuts were then applied and the decay
spectrum was assembled.  A muon hit was required in PC6, immediately
upstream of the stopping target, to ensure the muon reached the
target.  A hit in PC7, immediately downstream, vetoed the event.
Events with multiple muons were rejected.  The muon was required to
decay between 1~\us\ and 9~\us\ after the trigger; the delay ensured
that the ionization from the incident muon was collected before a hit
from the positron was recorded.  Events where a beam positron arrived
within 1~\us\ of either the muon or decay were rejected as well.
Additional cuts included the muon flight time through the M13 beam
line, used to reject muons from pion decays in flight, and a
requirement that the muon stopped within 2.5~cm of the detector axis,
which ensured that all decay positron tracks within the fiducial
region (Sect.~\ref{sec:fiducial}) were fully contained within the
detector.  Note that these cuts depend on observations of the muon
prior to decay or of the time of decay, and are hence unbiased in
terms of the properties of the positron.

\begin{figure}[!hbt]
  \includegraphics[width=3.4in]{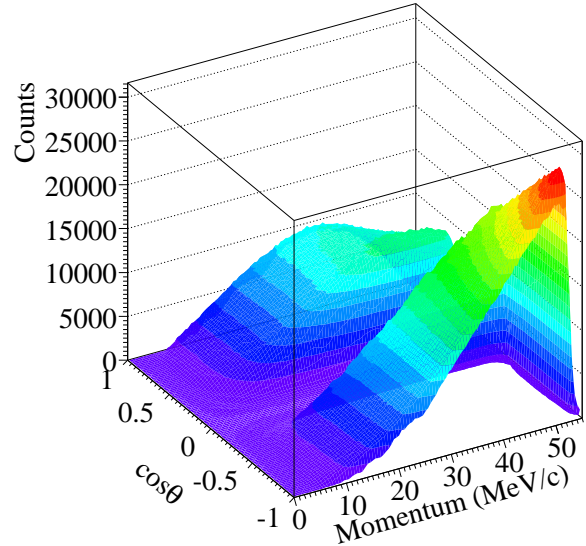}
  \caption{(Color online.)  Reconstructed muon decay spectrum.}
  \label{fig:dataspectrum}
\end{figure}

Figure~\ref{fig:dataspectrum} shows a muon decay spectrum
reconstructed from the ``Centred Stops'' data set (3\e{8} muon
decays; see Table~\ref{tab:datasets}).

\subsection{Energy Calibration}
\label{sec:ecal}

\begin{figure}[!hbt]
  \includegraphics[width=3.4in]{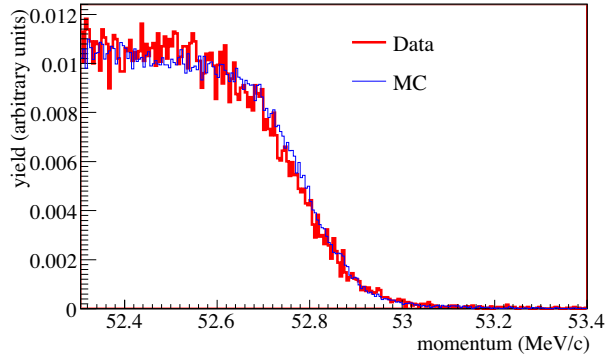}
  \caption{(Color online.)  Uncalibrated data and simulation histograms in the momentum range of the decay spectrum used for
    energy calibration, for $ -0.53 < \cos\theta < -0.50$.} 
  \label{fig:endpoint}
\end{figure}

Differences in the details of the muon stopping position within the
target foil, differences in the STRs, the accuracy of the magnetic
field maps, and knowledge of the wire positions resulted in
differences in the energy scales of the simulation and the data.  The
kinematic endpoint region (near $52.3<p<53.4$~\mevc) was used to
determine the relative energy calibration between the data and the
simulation after reconstruction was complete.  The relative positions
of the endpoint were compared between data and simulation in bins of
angle (see Fig.~\ref{fig:endpoint} for an example), and the
differences were parameterized separately for upstream and downstream
parts of the spectrum, according to
\begin{equation}
  \label{eq:ecal_endpoint}
  \Delta p_e = B_i + \frac{A_i}{\cos\theta},\quad i =
  \text{US, DS},
\end{equation}
where $\Delta p_e$ is the difference in the momentum of the spectrum
endpoint between data and simulation.  
Equation \ref{eq:ecal_endpoint} describes an  energy 
loss  proportional to the amount of material encountered by a
particle;  for a planar detector the amount of material is
proportional to $1/\cos\theta$. This form was found to describe
$\Delta p_e$ well, also.  Uncertainties in the calibration parameters
$B_i$ and $A_i$ were statistical.  This relative calibration technique
replaced the calibration using an analytic approximation to the end
point shape used in the previous analyses~\cite{musser:2005,
  gaponenko:2005, jamieson:2006}.

\subsection{Fiducial Region}
\label{sec:fiducial}

\begin{figure}[!hbt]
  \includegraphics[width=3.4in]{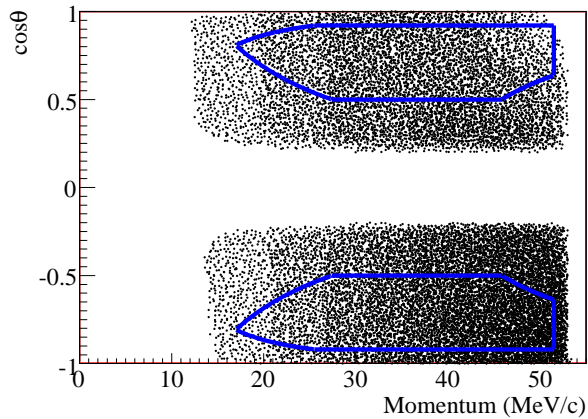}
  \caption{(Color online.)  A reconstructed muon decay spectrum; the
    dot density is proportional to the number of positron tracks
    reconstructed in a $(p,\cos\theta)$ bin.  The outlines show the
    fiducial region used for this analysis.}
  \label{fig:fiducial}
\end{figure}

Limitations of the reconstruction and physical aspects of the detector
required the imposition of fiducial cuts, which were applied after the
data and simulation were calibrated and assembled into spectra.
Improvements in reconstruction allowed us to use a larger fiducial
region for this analysis than was used for previous TWIST
measurements~\cite{musser:2005,gaponenko:2005,jamieson:2006}.  The
fiducial region adopted for this analysis required
$0.50<\abs{\cos\theta}<0.92$, $p<51.5$~\mevc, $10.0<p_t<39.7$~\mevc,
and $\abs{p_z}>13.7$~\mevc.  This region is shown in
Fig.~\ref{fig:fiducial}.

The 51.5~\mevc\ momentum cut removed the region of the spectrum used
by the energy calibration; this also removed the only sharp feature of
the spectrum, greatly reducing the sensitivity of the decay parameters
to the resolution.  The minimum $\abs{\cos\theta}$ cut eliminated high
angle events, which underwent large amounts of multiple scattering and
were difficult to reconstruct.  The maximum $\abs{\cos\theta}$ cut was
limited by increased reconstruction bias and a difference in
reconstruction efficiency between data and simulation at small angles.
The minimum $p_t$ cut eliminated tracks with radii too small to
provide sufficient transverse separation in the hits for reliable
fits.  The maximum $p_t$ cut, in combination with the requirement that
muon stopped within 2.5~cm of the detector axis, ensured that all
accepted decay positrons were fully contained in the detector.  The
minimum $\abs{p_z}$ cut removed tracks with wavelength comparable to a
detector periodicity of 12.4~cm.  The robustness of the analysis
against variations in these cuts was checked using variations of 0.02
in $\cos\theta$ or 0.5~\mevc\ for the momentum cuts; neither the decay
parameter measurements nor the systematic uncertainties were changed
by more than $1\e{-5}$.

The fiducial cuts were applied to the spectrum histogram, and were
therefore influenced by the histogram binning.  A histogram bin was
included in the fit if the $(p,\cos\theta)$ value of its centre passed
the above fiducial cuts.  The widths of the spectrum histogram bins
were 0.5~\mevc\ in momentum, and 0.02 in $\cos\theta$.

\subsection{Spectrum Fitting}
\label{sec:McFit}

The decay parameters were determined by comparing the shape of the
two-dimensional data spectrum to that of a simulated spectrum with
matching conditions.  The muon decay spectrum is linear in the
parameters $(\rho, \eta, \pmuxi, \pmuxi\delta)$, so the data spectrum
can be described in terms of the MC spectrum as
\begin{equation}
  \label{eq:mcfitter}
  \begin{split}
    S_D = S_M &+ \frac{\partial S}{\partial \rho}\Delta\rho +
    \frac{\partial S}{\partial \eta}\Delta\eta \\
    &+ \frac{\partial S}{\partial \pmuxi}\Delta(\pmuxi) +
    \frac{\partial S}{\partial \pmuxi\delta}\Delta (\pmuxi \delta),
  \end{split}
\end{equation}
where $S_D = d^2\Gamma_D/dx\,d(\cos\theta)$ is the decay spectrum from
data, and $S_M = d^2\Gamma_M/dx\,d(\cos\theta)$ is the spectrum
simulated using Eqn.~\eqref{eq:mudecay_michel} to generate the muon
decays.  Note that the simulated spectrum is generated using hidden
values of the decay parameters.  $\partial S/\partial\rho$ etc.\ are
the derivatives of Eqn.~\eqref{eq:mudecay_michel} with respect to the
decay parameters.  Radiative corrections were left out of these
derivatives, under the assumption that the dependence of the radiative
corrections on the decay parameters is negligible; the exception was
$\partial S/\partial\pmuxi$, which included the anisotropic radiative
corrections to facilitate the consistent treatment of \pmuxi\ as a
product.

``Derivative spectra'' were fully simulated and analyzed spectra in
the same way as $S_M$.  To generate these spectra, the magnitudes of
derivatives of Eqn.~\eqref{eq:mudecay_michel} with respect to the
decay parameters ($\abs{\partial S/\partial\rho}$ etc.) were treated
as probability distributions for the purposes of spectrum generation;
a sign was associated with each event according to the sign of the
derivative at that point, and this sign was used as a weight when the
reconstructed event was included in the final derivative spectrum.

The coefficients $\Delta\rho$, $\Delta(\pmuxi)$, and
$\Delta(\pmuxi\delta)$ are the fit parameters, and represent the
differences in the decay parameters between data and simulation.  The
muon decay spectrum in our fiducial does not accurately determine
$\eta$ because of the $x_0$ coefficient in Eqn.~\eqref{eq:Fis}, so the
value $\eta=-0.0036\pm 0.0069$ from the recent global
analysis~\cite{gagliardi:2005} was assumed.

Only differences between the real decay parameters in data and those
assumed by the simulation were measured directly in this procedure.
The values of the parameters assumed by the simulation ($\rho_h,
\delta_h, \xi_h$) were chosen randomly within 0.01 of the SM values
while being constrained to physically meaningful values (e.g.\
$\xi\delta/\rho \leq 1$), and these values were kept hidden throughout
the analysis and the evaluation of systematic uncertainties.
$\delta$, and hence $\Delta\delta$, was extracted from
$\Delta(\pmuxi\delta)$ as $(\xi_h\delta_h +
\Delta(\pmuxi\delta))/(\xi_h + \Delta(\pmuxi))$.  When measuring
systematic uncertainties, it was sufficient to assume Standard Model
values for the extraction of $\Delta\delta$.  In this way the
measurement outcome remained unknown until the values of the hidden
parameters were revealed.

\begin{figure}[!hbt]
  \centering
  \includegraphics[width=3.4in]{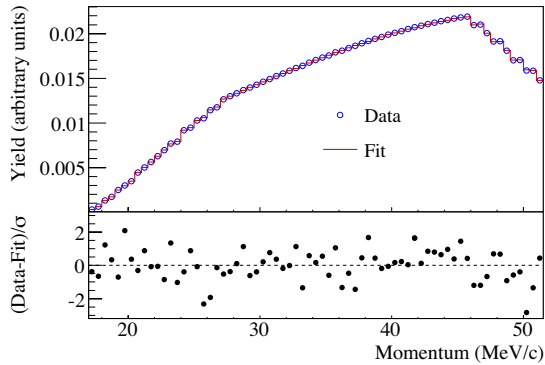}
  \caption{(Color online.) Momentum distribution of accepted events,
    ``Centred Stops'' data set.  Deviation from the shape of
    Eqn.~\eqref{eq:mudecay_michel} below 28~\mevc\ and above 45~\mevc\
    comes from the acceptance constraint shown in
    Fig.~\ref{fig:fiducial}.  The lower plot shows the difference
    between data and fit for each bin, normalized by the statistical uncertainty
    in the difference.}
  \label{fig:pspectoverlay}
\end{figure}

\begin{figure}[!hbt]
  \centering
  \includegraphics[width=3.4in]{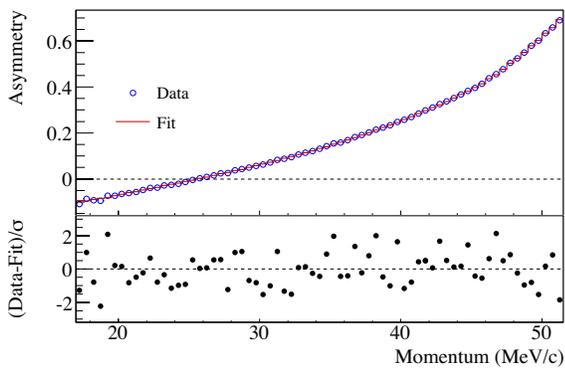}
  \caption{(Color online.) Decay asymmetry of accepted events,
    ``Centred Stops'' data set.  The lower plot shows the difference
    between data and fit for each bin, normalized by the statistical uncertainty
    in the difference.}
  \label{fig:asymvsp}
\end{figure}

The ability of the fitted Monte Carlo spectra to describe the data can
be viewed in terms two distributions: the angle-integrated momentum
distribution of the decay positrons, which is governed by $\rho$ (with
$\eta$), and the momentum dependence of the decay asymmetry, governed
by $\delta$.  The momentum spectrum of accepted events from a standard
data set is compared with that from the corresponding fitted
simulation in Fig.~\ref{fig:pspectoverlay}.  The decay asymmetry as a
function of momentum for accepted events from the same data set is
compared with the fitted simulation in Fig.~\ref{fig:asymvsp}.  Here
asymmetry is defined as $(B-F)/(B+F)$, where $B$ and $F$ are the
number of counts in the backward ($\cos\theta<0$) and forward
directions in a given momentum bin.  The normalized residuals
demonstrate the quality of the fit and the lack of bias.

This study used the same data analyzed by Jamieson et al.\ for the
direct measurement of the decay parameter \pmupixi, published in
2005~\cite{jamieson:2006}.  However, the \pmupixi\ analysis did not
include the study of the $\rho$ or $\delta$ systematic uncertainties
and corrections.  Furthermore, the present measurement used more
advanced analysis and simulation techniques than were available for
the previous study, as described below.  A new, independent set of
hidden decay parameters was generated for use in the simulation, and
this new set was used exclusively for this analysis.  Therefore, as
with the previous measurements from TWIST, the measurement outcome
could not be known until the hidden decay parameters were revealed.

\section{Substantial Changes Since Previous Studies}
\label{sec:improvements}

There were a number of improvements to the experimental apparatus and
data-taking techniques between 2002, when data were taken for the
first TWIST measurements of $\rho$ and $\delta$~\cite{musser:2005,
  gaponenko:2005}, and 2004, when the data were taken for both the
first TWIST measurement of \pmupixi~\cite{jamieson:2006} and the
present measurement of $\rho$ and $\delta$.  The geometry and material
of the muon stopping target used in 2004 was much better known than
the target used previously.  Moreover, additional monitoring and
feedback controls were implemented for the taking of data in 2004.
These improved the stability of the dipole magnets in the beamline,
the flatness of the chamber foils, average muon stop position, and
other aspects of the experiment, significantly increasing the quality
of the data.  The new muon beam monitoring chamber allowed
significantly better characterization of the muon beam for input into
the simulation, as well.

Since the previous measurements, the simulation and analysis were
modified to account for the asymmetric construction of the drift
chambers, where the wires were not centred between the foils (see
Sect.~\ref{sec:setup}).  The track reconstruction used by the present
analysis accounted for energy loss by the positrons, reducing
reconstruction bias by several \kevc; other reconstruction
improvements further reduced the reconstruction biases, particularly
at low angles.  Energy calibration used by the present analysis was
performed as a relative calibration between simulation and data,
resulting in a more accurate calibration.

As this analysis was nearing completion, a technique was developed to
extract STRs directly from the data.  This occurred too late to be
incorporated into this entire analysis, but it was used to estimate
the systematic uncertainty due to mismatch in the chamber response
between data and MC (see Sect.~\ref{sec:chamberresponse}).

\section{Simulation Validation}
\label{sec:usstops}

Since the data are fitted with simulated spectra to measure the muon
decay parameters, the simulation must correctly reproduce physical and
detector effects in an unbiased way; of particular importance are
energy loss, multiple scattering, and the probabilities for the
production of delta rays and bremsstrahlung.  Direct comparisons
between simulation and data were performed with specialized data not
used for decay parameter measurements.  In this mode, labeled
``upstream stops'', the momentum of the muon beam was reduced so that
the muons stopped upstream of the DCs.  Downstream-going decays then
passed through the entire detector, and were reconstructed twice,
first using only the upstream half of the detector, then using only
the downstream half.  Energy loss, scattering, helix fitter biases,
and reconstruction resolution all resulted in differences in the
properties of the two tracks.  Distributions of $\Delta p = p_{DS} -
p_{US}$, $\Delta\theta = \theta_{DS} - \theta_{US}$, and other
differences were used to examine reconstruction and physical effects,
independent of the shape of the decay spectrum, and to compare the
simulation directly to the data.

About $8\e{7}$ upstream stop events were considered for this analysis,
with $1.5\e{6}$ events accepted after cuts.  About ten times that
amount of simulated decays were generated.

\begin{figure}[!hbt]
    \includegraphics[width=3.4in]{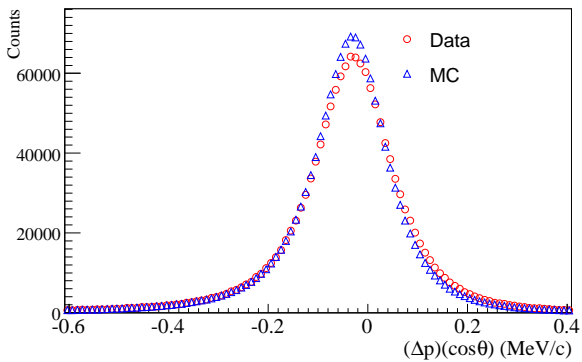}  
    \caption{(Color online.)  Energy loss distributions for positrons
      across the target region (Al and PCs), showing events with small
      losses and integrated over the fiducial region.}
  \label{fig:dpcosth_peak}
\end{figure}

\begin{figure}[!hbt]
    \includegraphics[width=3.4in]{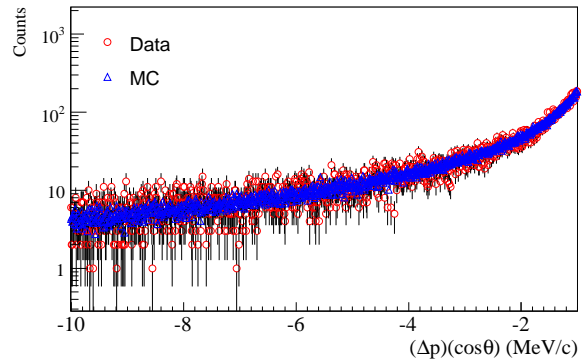}
    \caption{(Color online.)  As Fig.~\ref{fig:dpcosth_peak}, for
      large energy losses.}
  \label{fig:dpcosth_tail}
\end{figure}

As mentioned earlier, to first order energy loss is proportional to
the amount of material a particle encounters; in TWIST's planar
geometry this was proportional to $1/\cos\theta$.  Multiplying $\Delta
p$ by $\cos\theta$ removes this first-order angle dependence.
Figure~\ref{fig:dpcosth_peak} shows the distributions of $(\Delta
p)(\cos\theta)$, for both data and simulation; the simulation has been
normalized to data by integrated counts.  The most probable value was
$-28.4\pm 0.1$~\kevc\ for data and $-29.65\pm 0.04$~\kevc\ for MC; the
FWHM width was $155.9\pm 0.1$~\kevc\ for data and $141.64\pm
0.04$~\kevc\ for MC.  The slight difference in most probable energy
loss was compensated by the energy calibration.  The difference in the
width of the distribution leads to a systematic error in the
measurement of the decay parameters of 1.3\e{-4}, as discussed in
Sect.~\ref{sec:resolution}, and a correction was applied
(Table~\ref{tab:corrections}).
Figure~\ref{fig:dpcosth_tail} shows the same distributions but for
events where $(\Delta p)(\cos\theta) < -1$~\mevc, demonstrating very
good agreement between data and simulation over several orders of
magnitude.  This was used to determine part of the systematic
uncertainty due to the simulation of positron interactions
(Sect.~\ref{sec:positroninteractions}), which was found to be less
than 0.7\e{-4} for both $\rho$ and $\delta$.

\begin{figure}[!hbt]
    \includegraphics[width=3.4in]{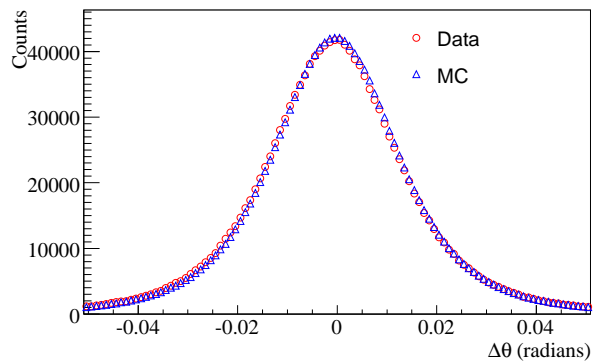}
    \caption{(Color online.)  Scattering angle distributions for
      positrons across the target region (Al and PCs), showing events
      with small angle scatters and integrated over the fiducial
      region.}
  \label{fig:dth_peak}
\end{figure}

\begin{figure}[!hbt]
    \includegraphics[width=3.4in]{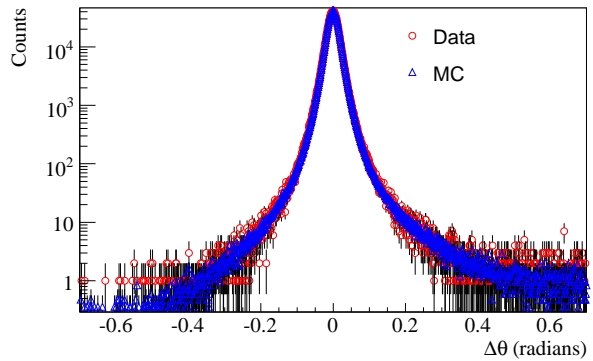}
  \caption{(Color online.)  As Fig.~\ref{fig:dth_peak}, for large
    angle scatters.}
  \label{fig:dth_tail}
\end{figure}

Figure~\ref{fig:dth_peak} shows the distributions of $\Delta\theta$
for data and simulation; the simulation was normalized to the data as
before.  The most probable value was $-0.97\pm 0.02$~mrad for data and
$-0.581\pm 0.007$~mrad for MC; the FWHM width was $29.75\pm 0.02$~mrad
for data and $29.159\pm 0.007$~mrad for MC.  The difference in the
most probable $\Delta\theta$ values between data and MC is very small
compared to the approximately 10~mrad angular resolution and so did
not affect the decay spectrum.  Unlike the energy loss, the slight
difference in the width of the $\Delta\theta$ distributions had a
negligible effect on the measurement of the decay parameters.
Figure~\ref{fig:dth_tail} shows the same distributions but including
events with large $\Delta\theta$, again demonstrating very good
agreement between data and simulation over many orders of magnitude.

The upstream stops technique was also used to compare reconstruction
efficiencies in data and simulation, as a function of momentum and
angle.  A simplified definition of reconstruction efficiency was used
to avoid introducing artifacts at the edge of the fiducial region and
other complications.  Let $r_d(p,\cos\theta)$ be the number of
downstream tracks in a given $(p,\cos\theta)$ bin for which an
upstream track was also reconstructed, and let $T_d(p,\cos\theta)$
represent the total number of downstream tracks in that bin.  Then the
upstream efficiency was defined as $\epsilon_u(p,\cos\theta) =
r_d/T_d$.  Downstream efficiency was similarly defined.

\begin{figure}[!hbt]
  \centering
  \includegraphics[width=3.4in]{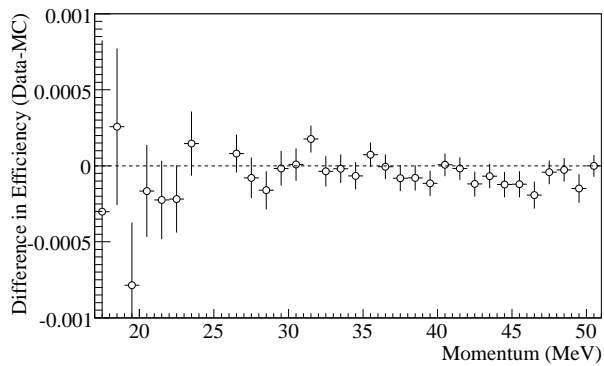}
  \caption{Difference between reconstruction efficiencies between data
    and simulation (MC) within the fiducial region as a function of
    momentum, for the upstream half of the detector.  The difference
    for the downstream half is similar.}
  \label{fig:ineffdiff_pproj}
\end{figure}

\begin{figure}[!hbt]
  \centering
  \includegraphics[width=3.4in]{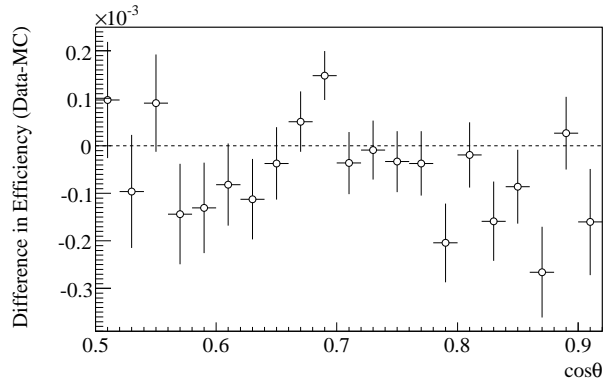}
  \caption{Difference between reconstruction efficiencies between data
    and simulation (MC) within the fiducial region as a function of
    $\cos\theta$, for the upstream half of the detector.  The
    difference for the downstream half is similar.}
  \label{fig:ineffdiff_cproj}
\end{figure}

Figures~\ref{fig:ineffdiff_pproj} and~\ref{fig:ineffdiff_cproj} show
the (Data--MC) differences in reconstruction efficiency, as functions
of momentum and $\cos\theta$ respectively; the differences of upstream
efficiencies are shown, and the downstream efficiencies were similar.
Each bin represents an average across the fiducial region.  The region
around the upstream stops beam momentum of $\sim 25$~\mevc\ has been
excluded to avoid an artifact in the efficiency measurement caused by
the beam positron phase space.  The simulation and data show good
agreement in reconstruction efficiency over the entire fiducial
region, and the slight differences are independent of momentum and
angle.  Overall, we find that all but $(5.7\pm 0.2)\e{-4}$ [$(5.12\pm
0.05)\e{-4}$] of the upstream stops events in the data [simulation]
that contain a reconstructed downstream track within the fiducial
region also contain a reconstructed track in the upstream half of the
detector; similarly, all but $(8.2\pm 0.2)\e{-4}$ [$(7.87\pm
0.06)\e{-4}$] of the events that contain a reconstructed upstream
track within the fiducial region also contain a reconstructed track in
the downstream half of the detector.  The differences in upstream and
downstream reconstruction efficiency give rise to a systematic error
of less than 0.6\e{-4}, as discussed in
Sect.~\ref{sec:chamberresponse}.

\section{Systematic Errors and Uncertainties}
\label{sec:systematics}

% Systematics and corrections, including summary table and discussion
% of bugs found.

\subsection{General Procedure}
\label{sec:systematics_procedure}

A systematic uncertainty is an uncertainty in the decay parameters as
the result of an unknown or unaccounted-for value or variation of some
experimental parameter.  A summary of the systematic uncertainties for
this measurement is given in Table~\ref{tab:sys_summary} by category;
these are described in more detail below.
\begin{table}[!hbt]
  \caption{Summary of systematic uncertainties by category.}
  \label{tab:sys_summary}
  \begin{tabular}{lcc}
    \hline
    \hline
    Category & $\Delta\rho$ & $\Delta\delta$ \\
    \hline
    Chamber Response & 0.00029 & 0.00052 \\
    Energy Scale & 0.00029 & 0.00041 \\
    Positron Interactions & 0.00016 & 0.00009 \\
    Resolution & 0.00002 & 0.00003 \\
    Alignment and Lengths & 0.00003 & 0.00003 \\
    Beam Intensity  & 0.00001 & 0.00002 \\[1ex]
    Correlations with $\eta$ & 0.00011 & 0.00001 \\
    Theory & 0.00003 & 0.00001 \\
    \hline
    Total & 0.00046 & 0.00067 \\
    \hline
    \hline
  \end{tabular}
\end{table}
A systematic error represents a known experimental bias rather than an
uncertainty; where these errors were found to be significant, a
correction was applied.  Very small errors were simply included in the
total systematic uncertainties.  Table~\ref{tab:corrections} lists the
corrections applied to the present measurement.
\begin{table}[!hbt]
  \caption{List of corrections to final decay parameter
    measurement, in units of $10^{-4}$.  Uncertainties in the
    corrections are included in the systematic uncertainties in
    Table~\ref{tab:sys_summary}.}
  \label{tab:corrections}
  \begin{tabular}{lcc}
    \hline
    \hline
    Correction & $\Delta\rho$ & $\Delta\delta$  \\
    \hline
    Drift time maps & $-2.0\pm 2.9$ & $+1.6\pm 5.2$  \\
    Momentum resolution & $+1.2\pm 0.2$ & $+1.3\pm 0.3$ \\
    \hline
    Total & $-0.8\pm 2.9$ & $+2.9\pm 5.2$ \\
    \hline
    \hline
  \end{tabular}
\end{table}

Several systematic uncertainties important to the measurement of
\pmupixi\ were not studied for this measurement.  The leading
systematic uncertainty in the previous measurement of \pmupixi\ by
TWIST~\cite{jamieson:2006} arose from uncertainty in our knowledge of
the muon beam emittance, and its impact on the muon depolarization as
the beam entered the magnetic field. The depolarization of the muons
due to the muon beam emittance was found to be 5--7\e{-3} depending on
the beam used, and the associated systematic uncertainty was
$\Delta\pmupixi = 0.0034$.  The muons depolarized further while at rest
in the stopping target at a rate of $(1.6\pm 0.3)\e{-3}$~\us$^{-1}$,
leading to an associated systematic uncertainty of $\Delta\pmupixi =
0.0012$.  We have no additional data to
improve our knowledge of the muon beam used at the time.  Furthermore,
whereas this issue had a very large impact on the determination of
\pmupixi, its impact on the measurement of $\rho$ and $\delta$ is
negligible.

Systematic uncertainties and errors were studied using the spectrum
fitting technique described by Eqn.~\eqref{eq:mcfitter}.  A new
simulated decay spectrum was produced by exaggerating an experimental
parameter at the simulation or analysis stage, and this was fit
against a simulation generated using the standard parameters.  The fit
results $\Delta\rho$ and $\Delta\delta$, scaled by the amount of
exaggeration, provided a direct measurement of the systematic
uncertainty associated with the experimental parameter.

All systematic errors and uncertainties were determined prior to
revealing the hidden decay parameters assumed by the simulation.

\subsection{Chamber Response}
\label{sec:chamberresponse}

To determine the effect of inaccuracies in the
\textsc{garfield}-generated STRs, analyses using STRs derived directly
from the data and the simulation were fit against the corresponding
standard analyses.  Data-derived and simulation-derived STRs for TWIST
have only recently been developed.  They were derived for data and
simulation from the means of the time residuals (the difference
between reconstructed hit time and measured hit time) as a function of
hit position within the drift cell, during reconstruction using
\textsc{garfield}-generated STRs.  These residuals were used to modify the
\textsc{garfield}-generated STRs, and the process was iterated until it
converged.  The data-derived STRs automatically account for variations
in temperature, foil positions, etc., within the detector, as well as
compensating for some residual biases in the helix reconstruction.
They lead to reduced $\chi^2$ values from the helix fitter.  They also
provide improved momentum resolution and a better match between data
and simulation compared to that described in Sect.~\ref{sec:usstops}.
Thus, we conclude that they are more correct for use in the analysis
than the STRs generated by \textsc{garfield}.

The tests determined by how much the measured decay parameters were
affected by the use of \textsc{garfield}-generated STRs in the
standard analysis.  The difference between the effect in simulation
and the effect in data demonstrated the amount by which the measured
decay parameters in a standard data-MC fit would shift; a correction
was applied to the final measurement to account for this
(Table~\ref{tab:corrections}).  The uncertainty on this difference, at
3\e{-4} for $\rho$ and 5\e{-4} for $\delta$, represented the
systematic uncertainty due to inaccuracies in the STRs.  This
dominated the chamber response uncertainties.

Other aspects of the chamber response studied included the
uncertainties in the DC foil positions, the asymmetries in the decay
reconstruction efficiency, and electronics-related wire time offsets.
None of these represented uncertainties greater than $0.6\e{-4}$ for
either $\rho$ or $\delta$, and most were much smaller.

\subsection{Energy Scale}
\label{sec:energyscale}

To translate uncertainties in the energy calibration parameters into
uncertainties in the decay parameters, new spectra were created by
applying energy calibration, with each calibration parameter
exaggerated, to a standard simulation.  These exaggerated spectra were
fit against a spectrum created using the standard calibration to
measure the effects on the decay parameters, which were found to be
3\e{-4} for $\rho$ and 4\e{-4} for $\delta$.  The systematic
uncertainty due to the energy calibration was the dominant uncertainty
related to energy scale.

Other aspects of the energy scale studied included the assumed
behavior of the energy calibration with momentum, and errors in the
shape of the magnetic field map used for analysis.  The energy
calibration was assumed to be independent of the positron momentum;
using a calibration proportional to the momentum changes the decay
parameters by less than $0.8\e{-4}$ for both $\rho$ and $\delta$.  The
simulated magnetic field used for analysis and simulation was compared
against the measured magnetic field map; the differences were found to
influence the decay parameters at the level of 0.7\e{-4} for both
$\rho$ and $\delta$.

\subsection{Positron Interactions}
\label{sec:positroninteractions}

\begin{figure}[!hbt]
  \centering
  \includegraphics[width=3.4in]{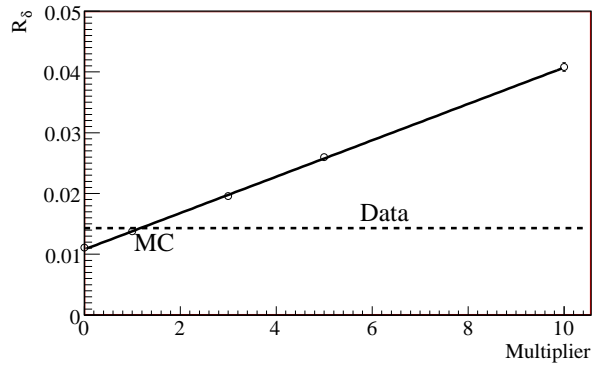}
  \caption{$R_\delta$ as a function of delta ray cross-section
    multiplier.  Open circles are simulation; standard simulation
    corresponds to a multiplicative factor of 1.  The linear fit was
    used to determine an effective multiplier for the data (dashed
    line).  Error bars are shown but are vanishingly small.}
  \label{fig:deltaboost}
\end{figure}

The production of delta rays and bremsstrahlung were the most
important discrete positron interactions for TWIST.  The simulation of
these was validated using the upstream muon stops described in
Sect.~\ref{sec:usstops} above.  The rate of delta ray production in
simulation was compared to that in data using the ratio
$R_\delta=N_{12}/N_{11}$, where $N_{12}$ is the number of events with
one upstream track and two downstream tracks, and $N_{11}$ the number
of events with one upstream track and one downstream track.  We find
$R_\delta = (1.432\pm 0.003)\e{-2}$ in the upstream stops data.
$R_\delta$ was also measured in several additional upstream stops
simulations with the delta ray production cross-section multiplied by
various factors, as shown in Fig.~\ref{fig:deltaboost}.
A linear fit to $R_\delta$ as a function of the cross-section
multiplier was used to translate the $R_\delta$ value from data into a
difference in delta ray cross-sections between data and simulation,
conservatively assuming that all events of this topology are due to
delta rays.  This is not true in practice, since $R_\delta$ does not
go to zero in the simulation when the delta ray production
cross-section in GEANT is set to zero (see Fig.~\ref{fig:deltaboost});
however, the conservative assumption was sufficient for estimating the
systematic uncertainty.  The effective delta ray production
cross-section determined from this method was found to be
approximately 18\% lower in the simulation than was seen in data.

\begin{figure}[!hbt]
  \centering
  \includegraphics[width=3.4in]{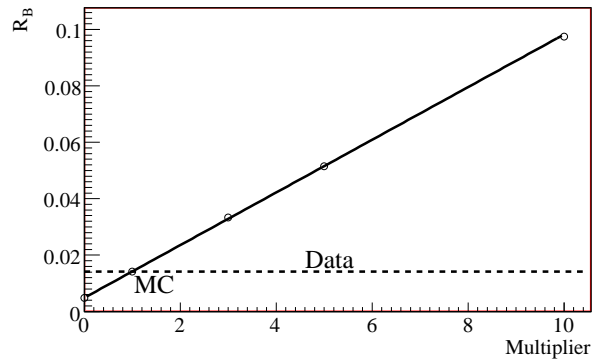}
  \caption{$R_B$ as a function of bremsstrahlung cross-section
    multiplier.  Open circles are simulation; standard simulation
    corresponds to a multiplicative factor of 1.  The linear fit was
    used to determine an effective multiplier for the data (dashed
    line).  Error bars are shown but are vanishingly small.}
  \label{fig:bremboost}
\end{figure}

The rate of bremsstrahlung production ($R_{B}$) in simulation was
compared to that in data by counting the number of through-going
positrons whose change in momentum was $p_{DS} - p_{US} < (\langle
\Delta p \rangle - 1$~\mevc), normalized to the total number of
reconstructed events; here $\langle\Delta p\rangle$ represents the
most probable value of $\Delta p$.  We find
$R_B = (1.42 \pm 0.01)\e{-2}$. As with the study of delta ray
production, $R_B$ was also measured in several additional upstream
stops simulations with the bremsstrahlung production cross-section
increased by various factors, as shown in Fig.~\ref{fig:bremboost}.
As before, a linear fit was used to translate the $R_B$ value from
data into a difference in bremsstrahlung cross-sections between data
and simulation; the simulation was found to agree with the data to
within 2\%.

The simulations produced with the delta ray or bremsstrahlung
cross-sections increased by a factor of three were fit against
standard simulation.  The systematic uncertainty due to errors in the
simulation of delta ray production was found to be 1.5\e{-4} for
$\rho$ and 0.9\e{-4} for $\delta$; that due to errors in the
simulation of bremsstrahlung production was less than 0.7\e{-4} for
either parameter.

Materials outside the sensitive region of the detector could scatter
particles back inside, resulting in extraneous hits that the track
reconstruction must sort out.  The number of backscattered positrons,
normalized by the number of muons, was studied in the data and the
simulation.  A new simulation was produced with an additional plate of
aluminum downstream of the detector as a backscattering source, and
this was fit against the standard simulation.  The effect on the decay
parameters was less than 0.4\e{-4}.

\subsection{Resolution}
\label{sec:resolution}

Reconstruction resolution in momentum and angle smeared the decay
spectrum.  The RMS angle resolution was about 5--15~mrad, and the
simulation agreed with the data to within 3~mrad.  The RMS momentum
resolution was 0.040--0.120~\mevc, and was smaller in the simulation by
0.007~\mevc.  Both types of reconstruction resolution varied with
energy and angle of the decay positron.  Since the spectrum was
inherently smooth, however, these resolutions did not significantly
distort the decay parameters directly.  The largest effect was at the
endpoint, where the momentum resolution changes the shape of the edge;
the 7~\kevc\ difference in resolution between simulation and data had
a small effect on the energy calibration, which resulted in a
distortion in the decay parameters.  Resolution was measured in the
data and MC as a function of momentum and angle using decays from
upstream muon stops, by measuring the widths of the energy loss and
scattering distributions.  New spectra were produced by smearing the
reconstructed momentum or angle of each event in the simulation before
including it in the spectrum and calculating the energy calibration.
The effect of the angle resolution on the decay parameters was found
to be less than 0.4\e{-4} for both $\rho$ and $\delta$.  The effect of
the momentum resolution on the decay parameters was 1.2\e{-4} for
$\rho$ and 1.3\e{-4} for $\delta$, and a correction was applied
(Table~\ref{tab:corrections}).

\subsection{Other Sources of Error}
\label{sec:othersystematics}

A number of other possible sources of systematic error were studied
and found to be small.

The effect of the uncertainty in the chamber alignment on the decay
parameters was tested by analyzing a simulation using distorted
chamber alignments, and fitting against the standard simulation.
These were found to affect the decay parameters by less than
0.03\e{-4}.  Errors in the transverse and longitudinal length scales
of the detector translated directly into errors on the transverse and
longitudinal components of the momentum; these were tested by
distorting these momentum components for reconstructed decays before
including them in a new spectrum.  The effect was less than 0.3\e{-4}.

New simulations were produced with exaggerated muon or positron beam
rates, to test the sensitivity of having an incorrect pile-up rate in
the simulation.  This showed the systematic uncertainty due to the
simulation of beam intensity to be less than 0.2\e{-4} for both beam
components.

As described above, the highest order of radiative corrections used
for this analysis was $O(\alpha^2 L)$.  The $O(\alpha^2)$ radiative
corrections~\cite{anastasiou:a2RCs} represented the theoretical error
in the $O(\alpha^2 L)$ corrections, and this determined the
theoretical uncertainty for this analysis.  A new simulation was
produced with the $O(\alpha^2 L)$ radiative corrections exaggerated,
and this was fit to the standard simulation.  The results show that
the theoretical uncertainty in the measured decay parameters is less
than 0.3\e{-4}.

During decay parameter fits, $\eta$ was normally fixed to the world
average value of -0.0036~\cite{gagliardi:2005}.  The standard fit
between data and MC was repeated with $\eta$ raised or lowered by one
standard deviation ($\delta\eta = \pm 0.0069$), giving the
correlations $\partial \rho/\partial\eta = 0.0162$,
$\partial\delta/\partial\eta = 0.0015$, and
$\partial\pmuxi/\partial\eta = 0.0155$.  This led to an uncertainty in
$\rho$ due to the uncertainty in $\eta$ of 0.00011.  Future
improvements in the knowledge of $\eta$ can be used to reduce this
systematic uncertainty directly.

Since this measurement was finalized, development of the simulation
and analysis software has continued for the analysis of additional
data.  During this process, an error in the particle identification
was found, which caused good events to be misclassified and discarded,
affecting about 0.6\% of events in the fiducial region.  This
distortion was primarily linear in $\abs{\cos\theta}$, which is a
shape not reflected in any of the derivative spectra; furthermore, the
distortion was the same in the analysis of both data and simulation,
leaving the derived decay parameters unaffected.  This software error
has been fixed for future analyses, but has been listed here for
completeness.

\section{Results}
\label{sec:results}

% McFit results for all sets, and averages; final rho & delta results
% after applying corrections and Black Box offsets; global analysis
% results.

The values of $\rho$ and $\delta$ measured from fitting each data set
with its corresponding simulation are listed in
Table~\ref{tab:McFits}.  After taking weighted averages and applying
the corrections in Table~\ref{tab:corrections}, we find
\[
\rho=0.75014\pm 0.00017(\text{stat})\pm 0.00044(\text{syst})\pm
0.00011(\eta),
\]
where the last uncertainty is due to the uncertainty
in $\eta$, and
\[
\delta = 0.75067\pm 0.00030(\text{stat})\pm 0.00067(\text{syst}).
\]
Both results are consistent with the Standard Model values of 3/4.
These represent a factor of two improvement over the previous TWIST
measurements~\cite{musser:2005, gaponenko:2005}.
% Values here are from the fixed-eta refits.
\begin{table}[!htb]
  \caption{Measured decay parameters from fits between data and
    simulation, for each data set. Uncertainties are statistical.
    Each fit has 2463 degrees of freedom.}
  \label{tab:McFits}
  \begin{tabular}{lccc}
    \hline
    \hline
    Set & $\rho$ & $\delta$ & $\chi^2$ \\
    \hline
    Mis-steered & $0.75054\pm 0.00044$ & $0.75066\pm 0.00077$ & 2505 \\
    PC5 stops & $0.75112\pm 0.00079$ & $0.74757\pm 0.00140$ & 2434 \\
    Stop $\frac{1}{2}$ & $0.74977\pm 0.00038$ & $0.75081\pm 0.00067$ & 2458 \\
    Stop $\frac{3}{4}$ A & $0.75024\pm 0.00037$ & $0.75073\pm 0.00066$ & 2483 \\
    High rate & $0.75003\pm 0.00050$ & $0.75047\pm 0.00088$ & 2392 \\
    Aperture & $0.75019\pm 0.00045$ & $0.75116\pm 0.00080$ & 2555 \\
    Stop $\frac{3}{4}$ B & $0.75042\pm 0.00049$ & $0.74882\pm 0.00086$ & 2468 \\
    \hline
    \hline
  \end{tabular}
\end{table}

The typical correlation coefficient between $\rho$ and $\delta$ from
the decay parameter fits is $+0.15$ for the fiducial range adopted
here.  Correlations also exist between the systematic uncertainties in
$\rho$ and $\delta$.  The contributions to $\sigma^2_{\rho\delta}$
associated with the chamber response and positron interaction
systematics are also positive and somewhat larger than the statistical
contribution from the decay parameter fits.  In contrast, the energy
calibration systematics for $\rho$ and $\delta$ are strongly
anti-correlated, and provide the dominant contribution to
$\sigma^2_{\rho\delta}$ in the present measurement.  Overall, we find
that the correlation coefficient between the total uncertainties in
$\rho$ and $\delta$ is $-0.16$.

The decay parameter fits also determined a value for \pmuxi, which can
be converted to a value for \pmupixi\ by correcting for known
depolarization effects.  After applying corrections corresponding to
the $\rho$ and $\delta$ corrections in Table~\ref{tab:corrections},
the fits give $\pmupixi=1.0025\pm 0.0004$(stat).  As explained in
Sect.~\ref{sec:systematics_procedure}, the systematic uncertainties
important to the measure of \pmupixi\ were not revisited during this
analysis, and this value should not be considered a revised
measurement.  When differences in the two analyses are considered, the
value of \pmupixi\ found here is consistent with the published TWIST
measurement of $\pmupixi = 1.0003\pm 0.0006(\text{stat})\pm
0.0038(\text{syst})$~\cite{jamieson:2006}.

% Art has pointed out that the STR systematic could legitimately be
% applied to the difference between the two analyses, but I don't know
% that it needs to be explicitly mentioned here.

\section{Global Analysis of Muon Decay}
\label{sec:globalanalysis}

In 2005 Gagliardi et al.\ performed a global analysis of all available
muon decay data~\cite{gagliardi:2005}, including earlier TWIST
measurements of $\rho$ and $\delta$~\cite{musser:2005,gaponenko:2005}.
That analysis has been repeated, incorporating the TWIST measurement
of \pmupixi~\cite{jamieson:2006} and the new $\rho$ and $\delta$
measurements presented here; the correlation factor of $-0.16$ between
the TWIST $\rho$ and $\delta$ measurements has also been included in the
calculation.  All other input values are the same as in the analysis
of Gagliardi et al.

In brief, the global analysis used a Monte Carlo method similar to
that of Burkard et al.~\cite{burkard:1985} to map out the joint
probability distributions for 9 independent variables, $Q_{RR}$,
$Q_{LR}$, $Q_{RL}$, $B_{LR}$, $B_{RL}$, $\alpha/A$, $\beta/A$,
$\alpha'/A$, and $\beta'/A$.  Each of these parameters is a bilinear
combination of the weak coupling constants $g_{\epsilon\mu}^\gamma$.
The decay parameters could then be written in terms of these
independent variables.  Table~\ref{tab:newglobalanalysis} shows the
results of this global analysis, as well as the results of the
analysis of Gagliardi et al.
\begin{table}[tbp]
  \caption{Results of a new global analysis of muon decay data,
    including the present measurements (90\% C.L.).
    $P_\mu^\pi=1$ is assumed.  Best fit values of selected decay
    parameters are also listed.}
  \label{tab:newglobalanalysis}
  \begin{tabular}{ccc}
    \hline
    \hline
    & Gagliardi et al.~\cite{gagliardi:2005} & Present Analysis \\
    & (\e{-3}) & (\e{-3}) \\
    \hline
    $Q_{RR}$ & $<1.14$ & $<0.96$ \\
    $Q_{LR}$ & $<1.94$ & $<1.38$ \\
    $Q_{RL}$ & $<44$ & $<42$ \\
    $Q_{LL}$ & $>955$ & $>955$ \\
    $B_{LR}$ & $<1.27$ & $<0.64$ \\
    $B_{RL}$ & $<10.9$ & $<10.8$ \\
    $\alpha/A$ & $0.3\pm 2.1$ & $0.1\pm 1.6$ \\
    $\beta/A$ & $2.0\pm 3.1$ & $2.1\pm 3.0$ \\
    $\alpha'/A$ & $-0.1\pm 2.2$ & $-0.1\pm 1.6$ \\
    $\beta'/A$ & $-0.8\pm 3.2$ & $-0.8\pm 3.1$ \\
    \hline
    \hline
    $\rho$ & $0.74959\pm 0.00063$ & $0.74964\pm 0.00035$ \\
    $\delta$ & $0.74870\pm 0.00114$ & $0.74997\pm 0.00065$ \\
    $\eta$ & $-0.0036\pm 0.0069$ & $-0.0042\pm 0.0064$ \\
    \hline
    \hline
  \end{tabular}
\end{table}

The 90\% confidence limits are given for the independent variables
listed above, and global best-fit values of the decay parameters
$\rho$, $\delta$, and $\eta$ are given.  The present analysis
represents significant improvements in the limits on $Q_{LR}$ and
$B_{LR}$, and tightens several of the other limits.  It is interesting
to note that the global analysis significantly reduces the uncertainty
in the value of $\rho$, from a total of 0.00063 to 0.00035.

The values of the $Q_{\epsilon\mu}$ from this global analysis can be
used in Eqn.~\eqref{eq:Qem} to place limits on the magnitudes of the
weak coupling constants $\abs{g_{\epsilon\mu}^\gamma}$; the exceptions
are $\abs{g_{LL}^V}$ and $\abs{g_{LL}^S}$, which are determined more
sensitively from inverse muon decay, {$e^-\nu_\mu \rightarrow \mu^-
  \nu_e$}.  The limits determined with this method are listed in
Table~\ref{tab:new_coupling_limits}, along with the values from the
previous global analysis~\cite{gagliardi:2005}.  The present analysis
represents a reduction of approximately 16\% in the limits for
$|g^S_{LR}|$ and $|g^T_{LR}|$, and a 30\% reduction for $|g^V_{LR}|$.
\begin{table}[tbp]
  \caption{Limits on the weak coupling constants.  (Limits on
    $|g_{LL}^S|$ and $|g_{LL}^V|$ are from Ref.~\cite{PDBook2006}.)}
  \label{tab:new_coupling_limits}
  \begin{tabular}{lcc}
    \hline
    \hline
    & Gagliardi et al.~\cite{gagliardi:2005} & Present Analysis \\
    \hline
    $|g_{RR}^S|$ & $< 0.067$ & $< 0.062$ \\
    $|g_{RR}^V|$ & $< 0.034$ & $< 0.031$ \\ 
    $|g_{LR}^S|$ & $< 0.088$ & $< 0.074$ \\
    $|g_{LR}^V|$ & $< 0.036$ & $< 0.025$ \\
    $|g_{LR}^T|$ & $< 0.025$ & $< 0.021$ \\
    $|g_{RL}^S|$ & $< 0.417$ & $< 0.412$ \\
    $|g_{RL}^V|$ & $< 0.104$ & $< 0.104$ \\
    $|g_{RL}^T|$ & $< 0.104$ & $< 0.103$ \\
    $|g_{LL}^S|$ & $< 0.550$ & $< 0.550$ \\
    $|g_{LL}^V|$ & $> 0.960$ & $> 0.960$ \\
    \hline
    \hline
  \end{tabular}
\end{table}

\section{Conclusion}
\label{sec:conclusion}

% Final rho & delta results; new QmuR limit; new LRS zeta limit;
% acknowledgements

This new measurement of the muon decay spectrum is a factor of two
more precise than previous measurements~\cite{musser:2005,
  gaponenko:2005}.  It is consistent with Standard Model predictions,
placing more stringent limits on ``new physics'' in the weak
interaction.  New indirect limits on the value of \pmupixi\ can be
obtained using these values of $\rho$ and $\delta$, in combination
with the the measurement of $\pmupixi\delta/\rho>0.99682$ at 90\%
confidence by Jodidio~\cite{Jodidio86:PmuXiDeltaRho,
  Jodidio86:erratum} and using the constraint $\xi\delta/\rho\leq 1$
(required to obtain positive definite decay probabilities from
Eqn.~\eqref{eq:mudecay_michel}).  Accounting for the correlation
coefficient of $-0.16$ between the $\rho$ and $\delta$ 
uncertainties, we find $0.99524 < P_\mu^\pi\xi \leq \xi < 1.00091$ at
90\% confidence.  This is a significant improvement over the previous
indirect limit of $0.9960 < P_\mu^\pi\xi \leq \xi <
1.0040$~\cite{gaponenko:2005}.

The quantity $Q_R^\mu = Q_{RR} + Q_{LR}$ represents the total
probability for a right-handed muon to decay into any type of
electron, a process forbidden under the Standard Model weak
interaction.  The new measurements of $\rho$ and $\delta$ lead to the
new limits on $Q_{RR}$ and $Q_{LR}$ shown in
Table~\ref{tab:newglobalanalysis}, and hence to a new 90\% confidence
limit upper bound on the combined probability $Q_R^\mu<0.0024$, a
slight improvement over the limit of $Q_R^\mu < 0.003$ from the
previous global analysis~\cite{gagliardi:2005}.

Under left-right symmetric models, $\rho>0.75$ is forbidden, so the
general measurement of $\rho$ can be converted into a 90\% confidence
limit lower bound within LRS models: $\rho>0.7493$ (compared with
$\rho > 0.7487$ from the $\rho$ measurement previously published by
TWIST).  The relation $\rho \simeq \frac{3}{4}(1-2\zeta_g^2)$ then
gives a 90\% confidence limit of $\abs{\zeta_g}<0.022$, a significant
improvement over the limit of $\abs{\zeta_g}<0.030$ for the previously
published TWIST value of $\rho$.

The final phase of TWIST is in progress.  Additional data are in hand,
taken in 2006 and 2007, and new analysis is underway.  Further
improvements to the simulation and analysis are being implemented,
including the use of measured drift time tables and improved track
reconstruction algorithms, and the resulting better agreement in the
momentum resolution between simulation and data.  These improvements
are expected to lead to additional factor-of-two reductions in the
uncertainties on $\rho$ and $\delta$, providing another incremental
improvement to searches for new physics.

We would like to thank C.A.~Ballard, S.~Chan, J.~Doornbos, B.~Evans,
M.~Goyette, D.~Maas, J.A.~Macdonald (deceased), T.A.~Porcelli,
N.L.~Rodning (deceased), J.~Schaapman, G.~Stinson, V.D.~Torokhov,
M.A.~Vasiliev, and the many undergraduate students who contributed to
the construction and operation of TWIST.  We also acknowledge many
contributions by other professional and technical staff members from
TRIUMF and collaborating institutions.  This work was supported in
part by the Natural Science and Engineering Research Council of Canada
and the National Research Council of Canada, the Russian Ministry of
Science, and the U.S.~Department of Energy.  Computing resources for
the analysis were provided by WestGrid.

% ***************** Bibliography **********************
% Paste the contents of the .bbl file here before submission!

\end{document}